\documentstyle[preprint,floats,tighten,prd,aps]{revtex}
\input epsf.tex
\def\desepsf(#1 width #2){\epsfxsize=#2 \epsfbox{#1}}
\begin{document}
\preprint{\vbox{
\hbox{OITS 640}
\hbox{October 1997}}  }
\draft

\title{Effective QCD coupling and power corrections \\
to photon-photon scattering}
\author{F.\ Hautmann}
\address{Institute of Theoretical Science, 
University of Oregon, Eugene OR 97403}

\maketitle

\begin{abstract}
 The scattering of two  off-shell photons 
is an infrared-safe process in QCD. For 
photon virtualities $Q$ in 
the range of a few ${\mbox {GeV}}$, accessible at LEPII, 
 power-behaved contributions  
 in $( \Lambda_{\rm {QCD}}/Q )^n$ 
to the total cross section 
may 
become non-negligible. Based on a 
 dispersion relation for  
the running coupling, 
we discuss   
these contributions  
and calculate the coefficients of the leading 
power correction  for transversely 
polarized 
and longitudinally polarized virtual photons. 
\end{abstract}


%****************************************************

%\narrowtext

Two-photon collisions provide one of the dominant modes in 
experiments at high-energy $e^\pm e^-$  
colliders. If the photons are sufficiently far off shell, the 
process is dominated by short-distance QCD interactions. A survey 
of the QCD studies  in photon-photon scattering 
that are currently being carried out 
at LEPII may be found in 
Ref.~\cite{miller}. In particular, 
the total cross section for scattering two 
off-shell photons is an infrared-safe observable. The role of this 
property has  recently been emphasized  in the context of  
 investigations of 
the high-energy limit of QCD~\cite{bhs}.  Predictions 
for $\sigma ( \gamma^* \, \gamma^*) $ 
 can be computed in perturbation theory and  
 tested as a function of the photon 
virtualities. For virtualities $Q$ 
of the order of a few ${\mbox {GeV}}$, 
 contributions to $\sigma$ suppressed 
by powers of $ \Lambda_{\rm {QCD}} / Q  $ 
may become non-negligible. These contributions are the subject of this 
paper.

A systematic approach to the calculation of power-behaved 
corrections to hard processes is 
 an open problem 
in QCD. For cases in which an 
operator product expansion is applicable, this provides a general 
framework to classify higher-twist contributions.  
However, so far  this has proved to be 
 of only limited practical use. 
 On the other hand, 
there have been efforts to develop methods that allow one to 
derive estimates of power corrections to a given observable 
from the study of the 
infrared behavior of its  perturbative series~\cite{muezakh}. 
In this context different techniques have been proposed 
over the past few years  and applied to a variety of hard 
processes (for recent reviews see, for instance, Ref.~\cite{bbtalk}). 
The basic observation underlying these methods is that  
the factorial growth of the 
coefficients of the QCD perturbation series 
in large orders 
gives rise to  ambiguities in the perturbative predictions 
which are  proportional to  power-behaved contributions. 
These ambiguities can be interpreted as being due to   an 
artificial separation between 
short-distance and long-distance physics in the 
perturbative treatment. 
  From the requirement 
that   
they must cancel in the 
physical cross sections once higher orders as well as  
nonperturbative contributions are  included,   
 one is able to  
  derive information on the structure of 
the power correction.

The physical origin of these power-like contributions 
 is an 
infrared one. They are 
associated  
with the loop integrations over 
 the regions of small momenta in Feynman graphs. 
Based on this, the authors of Refs.~\cite{dw95,dmw} 
have proposed a dispersion relation for the QCD running 
coupling in order to relate power-like corrections to 
the behavior of the coupling at small momentum scales.
In this paper, we will use this dispersion relation 
to analyze the exchange of gluons     in  
high-energy 
photon-photon interactions. 
 This  will enable us to identify the leading power correction 
to the total photon-photon cross section. 

\begin{figure}[htb]
\centerline{ \desepsf(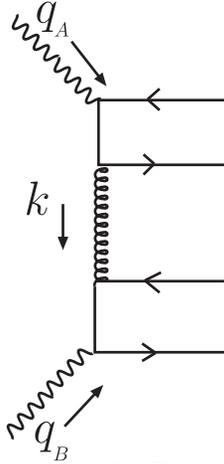 width 4 cm) } 
\caption{One of the graphs contributing to the Born amplitude for 
the process $\gamma^* + \gamma^* \to {\mbox {hadrons}}$ at large 
 $s$.   }
\label{figoneglu}
\end{figure}

We will 
discuss the scattering of two  off-shell (spacelike) photons    
with momenta $q_A $ and $q_B$, 
$ 
\gamma^*(q_A) + \gamma^*(q_B) \to {\mbox {hadrons}} 
$,  
with the 
virtualities $q_A^2 \equiv - Q^2_A$ and  
$q_B^2 \equiv - Q^2_B$  being large 
compared to  $\Lambda_{\rm {QCD}}^2 $.   
We will focus on the region where 
the center-of-mass energy 
$\sqrt{s} \equiv \sqrt{(q_A+q_B)^2}$  
is much larger than $Q_A$ and $Q_B$.   
In this region questions related to power-like  
corrections become especially important, because 
power-like corrections are expected to be associated 
with the mechanism that  unitarizes the total 
cross section at asymptotic energies.  
We will thus start with the 
large-$s$ 
form of the 
perturbative cross section, in which 
terms that fall like $Q^2 / s $ are neglected.   
In 
the Born approximation, 
 the  
 corresponding 
 amplitude 
is given by  graphs with  exchange of one gluon 
between  two quark-antiquark pairs created by the virtual photons 
(Fig.~\ref{figoneglu}).  In this approximation 
the total cross section has the structure~\cite{bhs} 
\begin{equation}
\label{convsigma} 
 \sigma_0 (s, Q_A^2, Q_B^2)
= 
{1 \over {2 \, \pi}} 
\,      
\int \, {{d^2 \, {\mbox{\bf k}}} \over { \pi}} \,
 {1 \over {({{\mbox{\bf k}}}^2)^2} } \,  
G({\mbox{\bf k}}^2, Q^2_A) \, 
G( {\mbox{\bf k}}^2, Q^2_B) \;\;\;\;.    
\end{equation} 
 Here $d^2 {{\mbox{\bf k}}}$ denotes  the 
integration over the transverse momentum flowing in the gluon line.  
The factors $1 / \left( {{\mbox{\bf k}}}^2 \right)^2$   
 come from the gluon propagators.
 The factors $G$ are each  
 proportional to $\alpha_s$, 
\begin{equation}
\label{gfunction} 
  G = \alpha_s \, g( {{\mbox{\bf k}}}^2 / Q^2 ) \;\;\; , 
\end{equation}
 and  
describe the 
coupling of the gluon to  the $q \, {\bar q}$ system.  
 The explicit form of the functions 
$g( {{\mbox{\bf k}}}^2 / Q^2 )$ depends on the photon polarization. 
  In what follows 
  we will 
  discuss first   
the cross section averaged over the two 
transverse  polarizations and then we will extend our results  
to the case of the longitudinal polarization.  

The cross section $\sigma_0$ has a 
scaling behavior in the 
photon virtualities 
 of the type 
$\sigma \propto 1/Q^2$, 
 modulated by 
logarithmic  
scaling-violation 
factors in the ratio of the two 
virtualities.    
 To evaluate the 
corrections to  
this behavior 
that are 
suppressed 
 by powers of the photon virtualities,    
 we  begin by introducing the running coupling 
$\alpha_s ( {\mbox{\bf k}}^2  ) $ 
for each one of the factors $G$ 
in the integral (\ref{convsigma}). Following the method of 
Refs.~\cite{dw95,dmw}, we 
 consider the  dispersive   representation  of the QCD coupling 
in terms of the spectral density $\rho_s$, 
\begin{equation}
\label{spectral} 
\alpha_s ( {\mbox{\bf k}}^2  ) 
 = -       
\int_0^{\infty} \, {{d \, \mu^2} \over { \mu^2 + {\mbox{\bf k}}^2}} \,
 \rho_s ( \mu^2  )\;\;\;\;,    
\end{equation} 
and we define the effective coupling $\alpha_{{\rm {eff}}}$ 
according to the relation  
\begin{equation}
\label{effdef} 
 \rho_s ( \mu^2  ) = { { \partial } \over { \partial \, \ln \mu^2} } 
\alpha_{{\rm {eff}}} ( \mu^2  )       
\;\;\;\;.     
\end{equation} 
We  thus have 
\begin{equation}
\label{alpharepr} 
\alpha_s ( {\mbox{\bf k}}^2  ) 
 =  {\mbox{\bf k}}^2 \,       
\int_0^{\infty} \, {{d \, \mu^2} \over 
{ (   {\mbox{\bf k}}^2 + \mu^2)^2}} \,
 \alpha_{{\rm {eff}}}  ( \mu^2  ) \;\;\;\;.     
\end{equation} 
The effective coupling $ \alpha_{{\rm {eff}}}$  
differs from the perturbative   
coupling in the infrared 
region.  The form of 
$ \alpha_{{\rm {eff}}}$ 
is determined by nonperturbative physics. As we shall see, the main 
point  
  of this 
  approach 
is 
 that, 
if 
this form can be assumed to be universal, then 
  power-behaved contributions 
to the cross section  
can be parametrized 
in terms of moments of the effective coupling variation  
at low transverse momentum scales.

 Replacing the 
strong coupling  in Eq.~(\ref{convsigma}) by
 Eq.~(\ref{alpharepr})  
yields     
\begin{equation}
\label{sigma1} 
 \sigma 
= 
{1 \over {2 \, \pi }} 
\,   
\int_0^{\infty} \, {{d \, \mu^2_A} \over { \mu^2_A}}\, 
\alpha_{{\rm {eff}}}  ( \mu^2_A  ) \, 
\int_0^{\infty} \, {{d \, \mu^2_B} \over { \mu^2_B}}\, 
\alpha_{{\rm {eff}}}  ( \mu^2_B  )
\, {\phi} ( Q_A^2, Q_B^2 ; \mu^2_A  , \mu^2_B ) 
\;\;\;\; ,    
\end{equation} 
 with the function ${\phi}$ being given by  
\begin{equation}
\label{phifun} 
{\phi} ( Q_A^2, Q_B^2 ; \mu^2_A  , \mu^2_B ) = \mu^2_A \, \mu^2_B \, 
\int_0^{\infty} \, {{d \, {{\mbox{\bf k}}}^2} \over 
{ (   {\mbox{\bf k}}^2 + \mu^2_A)^2 \; 
(   {\mbox{\bf k}}^2 + \mu^2_B)^2}} \,
\, g ( {{\mbox{\bf k}}}^2 / {Q}^2_A ) 
\, g ( {{\mbox{\bf k}}}^2 / {Q}^2_B )
\;\;\;\;.     
\end{equation} 
Eq.~(\ref{sigma1}) is written in terms of two dispersion variables 
$\mu_A$, $\mu_B$, 
corresponding to the 
fact that our starting amplitude  (Fig.~\ref{figoneglu}) 
is 
a two-loop contribution in QCD.  
 Introducing the variation 
$\delta  \alpha_{{\rm {eff}}} $ of the strong coupling 
at low scales,  we decompose $\alpha_{{\rm {eff}}}$ as follows  
\begin{equation}
\label{deltaeff} 
\alpha_{{\rm {eff}}} (\mu^2) = \alpha_{\rm {PT}} (\mu^2) +
\delta  \alpha_{{\rm {eff}}} (\mu^2) \hspace*{0.5 cm} , 
\end{equation}
where  $ \alpha_{\rm {PT}}$ denotes the form of the 
coupling  in perturbation theory. 
The decomposition (\ref{deltaeff}) gives rise to terms 
in  $\alpha_{\rm {PT}} \times \alpha_{\rm {PT}}$, 
$ \alpha_{\rm {PT}} \times \delta  \alpha_{{\rm {eff}}}$, and 
$\delta  \alpha_{{\rm {eff}}} \times \delta  \alpha_{{\rm {eff}}}$ 
in Eq.~(\ref{sigma1}).

The integration of the terms in 
$\alpha_{\rm {PT}} \times \alpha_{\rm {PT}}$
is  
dominated
by values of the dispersion variables of the order of the hard scales, 
$Q_A$ and $Q_B$, as we will see below. Thus, these 
terms give rise to the standard leading result for the cross section 
in perturbation theory. 
The 
terms involving $\delta  \alpha_{{\rm {eff}}}$, on the other hand, 
  probe the behavior 
of the function $\phi$ 
at small values of the 
dispersion variables, because 
the distribution of $\delta  \alpha_{{\rm {eff}}}$ 
 is 
concentrated at small scales. These terms 
are responsible for power-behaved corrections to  $\sigma$. 
 We will see that in the case of the leading power correction 
the quadratic term in $\delta  \alpha_{{\rm {eff}}}$ does not 
contribute and the correction comes entirely from the 
linear term in $\delta  \alpha_{{\rm {eff}}}$.

To study  
the form of the function $\phi$ at small  
$\mu^2$, we  
  use   
 the explicit expression for  $g$.  
For   
transversely 
polarized photons, this  is given by~\cite{bhs}    
\begin{equation}
\label{gtransv}
g ({{\mbox{\bf k}}}^2 / Q^2)  =
8 \, \alpha  \, \left( \sum_q e^2_q \right)  \, 
\int_0^1  \,
{d \, z} \,  P(z) \, \int_0^1  \,
{d \, \lambda}  \,  P (\lambda) \,
 { { \,  {{\mbox{\bf k}}}^2
 \,  
} \over {
 \lambda \, (1 - \lambda) \,{{\mbox{\bf k}}}^2 + z \, (1-z) \, Q^2
 } } \hspace*{0.5 cm} , 
\end{equation} 
where 
 $\alpha$ is the electromagnetic fine structure,  $e_q$ is the 
quark electric charge in units of $e = \sqrt{ 4 \, \pi \, \alpha}$, and  
$P$ is the 
vector $ \to $ fermion 
splitting function 
\begin{equation}
\label{px}
P(x) =  {{x^2 + (1 - x)^2} \over 2 } \hspace*{0.5 cm} . 
\end{equation}
From Eqs.~(\ref{phifun}) and 
(\ref{gtransv}) we may observe that 
for small   $\mu$ the function $\phi$ vanishes. 
This is associated with   
 the fact that the $\gamma^* \, \gamma^*$ 
cross section is an infrared safe quantity  
in perturbation theory  (see 
Eq.~(\ref{convsigma})), that is, 
 the factors 
$g$ vanish  like ${{\mbox{\bf k}}}^2 / Q^2$ (times logarithms) as 
${{\mbox{\bf k}}}^2 \to 0$. For large $\mu$  the function $\phi$ 
also vanishes. This is associated with the fact that  
 the graph in Fig.~\ref{figoneglu} is well-behaved in the 
ultraviolet,  
that is,  the factors $g$ do not 
grow more than logarithmically at large ${{\mbox{\bf k}}}^2$. 

\begin{figure}[htb]
\centerline{ \desepsf(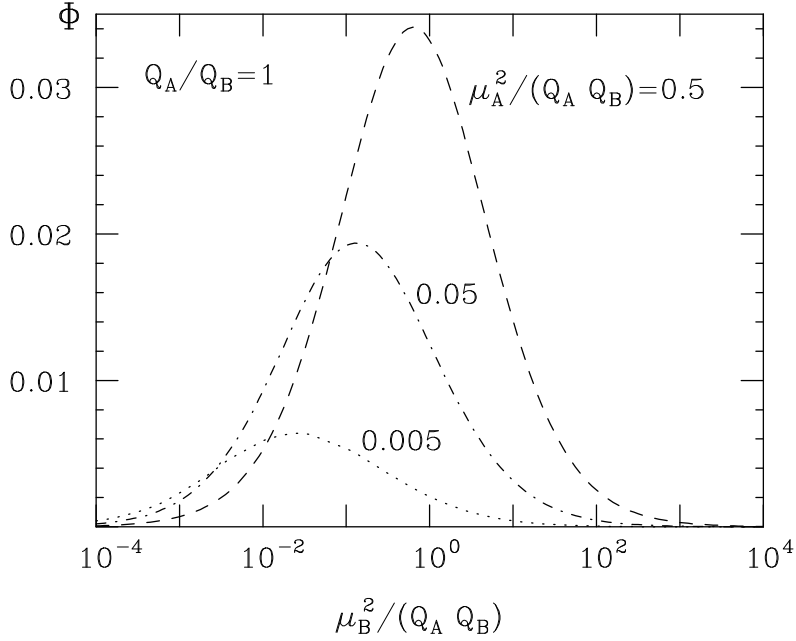 width 13 cm) } 
\caption{The dependence of the function $\phi$  
  on one of the 
dispersion variables ($\mu_B$) for different 
 values of the other one 
($\mu_A$). We take $Q_A = Q_B$ 
and plot the rescaled function $\Phi = Q_A \, Q_B\, \phi / ( 
8 \, \alpha \, \sum_q e_q^2 )^2$ versus $\mu_B^2/(Q_A \, Q_B)$. }
\label{figphimub}
\end{figure}

   We  now evaluate the function $\phi$ numerically. 
  In Fig.~\ref{figphimub} we 
plot the $\mu_B$-dependence 
of the dimensionless function $Q_A \, Q_B \, \phi$ 
 for fixed  (small) values of 
$\mu_A$, working at $Q_A = Q_B$. 
 First,  
we note   
that 
$\phi$ peaks in the region where the ratio $\mu_A / \mu_B$ of 
the dispersion variables is of order $1$ and their product 
is of order $Q_A \, Q_B$. Then the 
leading contribution to the term  
$\alpha_{\rm {PT}} \times \alpha_{\rm {PT}}$ 
from the integral (\ref{sigma1}) is obtained by 
pulling out of the integral the factors of 
$\alpha_{\rm {PT}}$ evaluated at a scale of 
order $Q_A \, Q_B$. 
We thus recover the leading perturbative result 
(\ref{convsigma}). 
Second, 
 we consider the term 
$ \delta  \alpha_{{\rm {eff}}} ( \mu_A^2) 
\, \alpha_{\rm {PT}} ( \mu_B^2)  $. 
We see from Fig.~\ref{figphimub}  
that for small $\mu_A$
the function $\phi$  
peaks   at a value 
${\overline \mu}_B^2$    
proportional to $\sqrt{Q_A \, Q_B}$ and to 
$\sqrt{\mu_A^2}$. 
 Then we  
approximately  
perform 
 the $\mu_B^2$-integration  
 by pulling  
out of the integral 
the factor of $\alpha_{\rm {PT}}$  
evaluated at the scale ${\overline \mu}_B^2$  and computing the 
integral of   
  $\phi$.  
By adding the symmetric term with  $\mu_A$ and $\mu_B$ interchanged,    
 we  write the contribution 
$\sigma_1$ 
from 
the terms of the first order 
in  $ \delta  \alpha_{{\rm {eff}}}$ 
  as 
\begin{equation}
\label{sigmax} 
 \sigma_{1}  
\simeq  
{1 \over { \pi }} 
\,    
\int_0^{\infty} \, {{d \, \mu^2_A} \over { \mu^2_A}}\, 
\delta \alpha_{{\rm {eff}}}  ( \mu^2_A  ) \; 
 \alpha_{{\rm {PT}}}  \left( Q_A \, Q_B \, \sqrt{\mu^2_A/(Q_A \, Q_B)}  
\right)
\; {\widetilde \phi} ( Q_A^2, Q_B^2 ; \mu^2_A   ) 
\;\;\;\; ,    
\end{equation} 
where  the function ${\widetilde \phi}$ 
is  defined as   
\begin{equation}
\label{phi1}
{\widetilde \phi} = 
\int_0^{\infty} \, {{d \, \mu^2_B} \over { \mu^2_B}}\,
\phi 
\;\;\;\; .       
\end{equation}

Next we use the fact 
 that the  requirement of 
 consistency with the operator product expansion~\cite{shif} 
   constrains   the structure 
of the effective coupling variation 
 $\delta \alpha_{{\rm {eff}}}$~\cite{dmw,bbz}. 
Let us define the moments 
\begin{equation}
\label{moments} 
A^{(n)}_{2 \, p} = {1 \over {2 \, \pi }} \, { { d^n } \over 
{d p^n} }  
\,   \int_0^{\infty} \, { {d  \, \mu^2} \over \mu^2} 
\, \left( \mu^2 \right)^p \,  
\delta \alpha_{{\rm {eff}}}  ( \mu^2  ) \, 
\;\;\;.   
\end{equation}  
  In order that the ultraviolet  behavior  of the running coupling 
  be not 
 ruined by the variation   $\delta \alpha_{{\rm {eff}}}$, 
the moments  with $n = 0$ and 
 $p$ integer have to 
vanish~\cite{dmw,bbz,bbb}. This implies that 
 only  terms in ${\widetilde \phi}$ that are nonanalytic in 
$\mu_A^2 $ 
for small $\mu_A^2 $ 
can 
contribute to  power corrections.  
 To find these terms, we examine the dominant region of 
integration 
  in 
Eq.~(\ref{phifun}),  
  $\mu_A^2 
{\raisebox{-.4ex}{\rlap{$\,\sim\,$}} \raisebox{.4ex}{$\,<\,$}} 
 {\bf k}^2 
{\raisebox{-.4ex}{\rlap{$\,\sim\,$}} \raisebox{.4ex}{$\,<\,$}} 
 Q_A \, Q_B$.  
By evaluating 
 the contribution  from this region,  
 we  
   obtain the following approximate expression 
for the function ${\widetilde \phi}$:   
\begin{eqnarray}
\label{phi1bis}
&& {\widetilde \phi} (Q_A^2, Q_B^2 ; 
\mu^2_A ) \simeq  \left( 8 \, \alpha \, 
 \sum_q e_q^2 \right)^2 \, {1 \over {Q_A \, Q_B}} \, ( - \eta ) \, 
\left\{  {2 \over 9} \, {\mbox {Li}}_3 (- {1 \over \eta}) + 
{1 \over 27} \,  {\mbox {Li}}_2 (- {1 \over \eta}) 
\right. 
\\
&& \left.
+
\left[ {35 \over 324} + {1 \over 36} \, \ln^2 \left( {Q_A^2 \over Q_B^2} 
\right) \right] \, \ln \left( { { 1+ \eta } \over \eta } \right) 
 +   
\left[ {49 \over 324} - {1 \over 36} 
\, \ln^2 \left( {Q_A^2 \over Q_B^2} 
\right) \right] \, {1 \over { 1+ \eta }} 
\right\}
\hspace*{0.2 cm} ,  \hspace*{0.4 cm} \eta \equiv  { \mu^2_A  \over 
{Q_A \, Q_B} } 
\hspace*{0.2 cm} , 
\nonumber  
\end{eqnarray} 
where  
\begin{equation} 
\label{dilog}
{\mbox {Li}}_2 (z) = - \int_0^z \, { {d t } \over t} \, \ln ( 1 - t ) 
\hspace*{0.6 cm} , \hspace*{1.2 cm} 
{\mbox {Li}}_3 (z) =  \int_0^z \, { {d t } \over t} \, 
{\mbox {Li}}_2 (t)
\hspace*{0.6 cm} . 
\end{equation}
 This formula is accurate up to the leading power 
in 
$\eta = \mu^2_A / (Q_A \, Q_B) $  
 and up to  the single logarithms.    
Higher powers in $\eta$ as well as 
constants  associated with  the leading power are changed  by 
the terms 
that we have dropped in 
performing  
the integral  (\ref{phifun}).   
To determine the 
nonanalytic behavior 
 that controls the leading power correction to $\sigma$   
 we expand around $\eta = 0$:  
\begin{eqnarray}
\label{trilog}
&& {\widetilde \phi} (Q_A^2, Q_B^2 ; 
 \mu^2_A 
 ) 
\simeq   \left( 8 \, \alpha \, 
 \sum_q e_q^2 \right)^2 \, {1 \over {Q_A \, Q_B}} \, {1 \over 27} \, 
\eta \, 
\left\{ \ln^3 {1 \over \eta} + {1 \over 2} \, \ln^2 {1 \over \eta} 
\right. 
\nonumber\\
&& + \left. 
\left[ 6 \, \zeta (2) - {35 \over 12} - {3 \over 4} \, 
\ln^2 \left( {Q_A^2 \over Q_B^2} 
\right) \right] \, \ln {1 \over \eta} 
+ \{ {\mbox {analytic}} \; {\mbox {terms}} \} \right\} 
+ {\cal O} (\eta^2) 
\hspace*{0.2 cm} ,  \hspace*{0.2 cm} 
\zeta (2) \simeq 1.64   \hspace*{0.1 cm} . 
\end{eqnarray} 	
Thus, from Eqs.~(\ref{sigmax}) and (\ref{moments})  we find 
\begin{equation}
\label{sigcr} 
\sigma_{1}  
\simeq  
{{ 
\left( 8 \, \alpha\,\sum e_q^2 \right)^2 \, 2 \, 
\alpha_{\rm {PT}}(Q_A Q_B)} \over { 27 \, Q_A^2 \, Q_B^2}} 
\,\left[-(3L^2+L+k)A^{(1)}_2+\left(3 \, L+{1\over 2}\right)
A^{(2)}_2   -A^{(3)}_2\right]
\end{equation}
where
\begin{equation}
\label{Lk} 
L=\ln(Q_A Q_B)\;\;,
\;\;\;\;\; 
k=
6 \, \zeta (2) - {35 \over 12} - {3 \over 4} \, 
\ln^2 \left( {Q_A^2 \over Q_B^2} \right)\;\;.
\end{equation}
Here we have written 
$  
\alpha_{\rm {PT}}(Q_A Q_B\sqrt{\eta})
=\alpha_{\rm {PT}}(Q_A Q_B) + {\cal O}(\alpha^2_{\rm {PT}}\,\ln\eta) 
$ 
and neglected the higher order term on the right-hand side. In the 
definition of $L$ in Eq.~(\ref{Lk}) we have implicitly taken the 
scale in the logarithm to be $1 \, {\mbox {GeV}}^2$.

Eq.~(\ref{sigcr}) gives the result for the 
leading 
power correction 
to the $\gamma^* \, \gamma^*$ cross section in terms of the 
dimensionful nonperturbative parameters $A_2^{(i)}$. These 
parameters are thought of as being universal and should be 
determined by fits to experimental data. 
The coefficients of these parameters   are given in 
Eq.~(\ref{sigcr})  
 up to terms that   vanish like   
  $  Q^2 / s$ at large $s$.  We see that  
the contributions 
 in $A^{(1)}_2$
and $A^{(2)}_2$ are enhanced by  
 double and single logarithms of 
the hard scale $Q_A \, Q_B$.

So far, there have 
been attempts to
 study  
 the 
nonperturbative moments 
$A^{(i)}_2$ 
 based on data 
 for   
deeply  inelastic
lepton-nucleon scattering 
and  for  
hadronic final states in $e^+ \, e^-$ annihilation.   
     Data on the structure function  $F_2$ 
 suggest that $A^{(1)}_2 \, 
\simeq -0.15 \, {\mbox {GeV}}^2$~\cite{dasg}. 
Very little is known about  
 $A^{(2)}_2$
and $A^{(3)}_2$ at present. 
$A^{(2)}_2$ enters in the power correction to the 
mean value of the three-jet resolution 
in $e^+ \, e^-$ annihilation. 
 A recent analysis of data recorded at PETRA 
 suggests that this correction 
 should be very small~\cite{rwthjade}. 
 Assuming that the 
contribution of $A^{(1)}_2$ predominates in  
Eq.~(\ref{sigcr}), owing to its 
enhanced coefficient, we expect the leading power correction 
to be positive.

 To complete our analysis, 
we need to show that, as anticipated,   
 the term  
of the second order in $ \delta  \alpha_{{\rm {eff}}}$ 
in Eq.~(\ref{sigma1})
does not contribute to the leading power correction. We write 
this term as   
\begin{equation}
\label{sigmadd} 
 \sigma_{2}  
= 
{1 \over {2 \, \pi }} 
\,   
\int_0^{\infty} \, {{d \, \mu^2_A} \over { \mu^2_A}}\, 
\delta \alpha_{{\rm {eff}}}  ( \mu^2_A  ) \, 
\int_0^{\infty} \, {{d \, \mu^2_B} \over { \mu^2_B}}\, 
\delta \alpha_{{\rm {eff}}}  ( \mu^2_B  )
\, {\phi} 
\;\;\;\; .    
\end{equation} 
 This contribution  probes the function $\phi$  in the corner of 
the phase space  where both 
$\mu^2_A$  and $ \mu^2_B $ are small. We  may   
restrict the ${{\mbox{\bf k}}}^2$-integration  
in Eq.~(\ref{phifun}) 
to the region    
 ${{\mbox{\bf k}}}^2 
{\raisebox{-.4ex}{\rlap{$\,\sim\,$}} \raisebox{.4ex}{$\,<\,$}} 
Q^2$, with 
$Q^2$ being of the order of the hard scales $Q_A^2$, $Q_B^2$, and  
we 
 may 
rewrite the function $\phi$ as 
 a sum of two pieces, each 
proportional to a logarithmic derivative with respect to one of the 
dispersion variables: 
\begin{eqnarray}
\label{deriv} 
{\phi} ( Q_A^2, Q_B^2 ; \mu^2_A  , \mu^2_B ) \simeq
&&  - {\partial \over {\partial \ln  \mu^2_A }} \left[ 
 { \mu^2_B \over { (\mu^2_A- \mu^2_B)^2} } \, 
\int_0^{Q^2} \, {{d \, {{\mbox{\bf k}}}^2} \over 
{    {\mbox{\bf k}}^2 + \mu^2_A }} \,
\, g ( {{\mbox{\bf k}}}^2 / {Q}^2_A ) \, 
g ( {{\mbox{\bf k}}}^2 / {Q}^2_B ) \right] 
\nonumber\\
&&  - {\partial \over {\partial \ln  \mu^2_B }} \left[ 
 { \mu^2_A \over { (\mu^2_A- \mu^2_B)^2} } \, 
\int_0^{Q^2} \, {{d \, {{\mbox{\bf k}}}^2} \over 
{    {\mbox{\bf k}}^2 + \mu^2_B }} \,
\, g ( {{\mbox{\bf k}}}^2 / {Q}^2_A ) 
\, g ( {{\mbox{\bf k}}}^2 / {Q}^2_B ) \right]
\;.     
\end{eqnarray} 
The pole at $\mu^2_A = \mu^2_B$ cancels in the sum of the two pieces. 
 This may be 
checked, for instance, by substituting the 
 expression  for $g$, expanding it for small ${\mbox{\bf k}}^2$  
and performing the integrals explicitly. Taking this cancellation 
into account, the first piece gives rise to terms that are analytic 
in $\mu^2_B$ and possibly  have nonanalytic  contributions in 
$\mu^2_A$. Analogously, the second piece 
gives rise to terms that are analytic 
in $\mu^2_A$ and possibly  have nonanalytic contributions in 
$\mu^2_B$. Since, as already noted, integer moments 
of $\delta \alpha_{{\rm {eff}}}$  
vanish, for each term we get 
  a vanishing contribution  
in Eq.~(\ref{sigmadd}) 
from either the integral in  $\mu^2_A$ or 
the integral in  $\mu^2_B$.  
Therefore, $\sigma_{2}$ does not 
contribute to the leading power correction.

Finally, 
let us 
 consider the 
scattering of longitudinally polarized photons. 
The function $g^{(L)}$ 
for this case 
has a structure analogous 
to Eq.~(\ref{gtransv}) but with a different splitting 
function, $P^{(L)}$~\cite{bhs}. 
The longitudinal splitting function 
 vanishes at the endpoints: 
\begin{equation}
\label{splilong}
 P^{(L)} = \sqrt{2} \,  x \, (1-x) 
\hspace*{0.4 cm} .     
\end{equation} 
This can be seen as being   
  associated with the absence of 
aligned-jet terms for longitudinal photon scattering at   
high energies~\cite{alj}. 
 As a result, by 
performing a calculation 
analogous to the one described above for the transverse case, 
we find that 
the  
small-$\eta$ expansion of  
 ${\widetilde \phi}^{(L)}$ 
 has at most single 
logarithms:  
\begin{equation}
\label{phi1long}
 {\widetilde \phi}^{(L)} (Q_A^2, Q_B^2 ; 
 \eta  
) 
\simeq   \left( 8 \, \alpha \, 
 \sum_q e_q^2 \right)^2 \, {1 \over {Q_A \, Q_B}} \, {1 \over 9} \, 
\eta \, \left[
 \ln {1 \over \eta} 
+ \{ {\mbox {analytic}} \; {\mbox {terms}} \} \right]  
+ {\cal O} (\eta^2) 
\hspace*{0.6 cm} .     
\end{equation} 
Therefore,  for the power correction to the 
longitudinal $\gamma^* \, \gamma^* $ cross section 
 $\sigma^{(L)}$  
 we get 
\begin{equation}
\label{sigcrlong} 
\sigma_{1}^{(L)}   
\simeq  
{{ 
\left( 8 \, \alpha\,\sum e_q^2 \right)^2 \, 2 
\, \alpha_{\rm {PT}}(Q_A Q_B)} \over { 9 \, Q_A^2 \, Q_B^2}} 
\, \left( -  \, A^{(1)}_2 \right) \hspace*{0.6 cm} . 
\end{equation}
We observe that 
the structure of the power correction is considerably more complicated 
in the transverse case. 
In the longitudinal case, the 
power correction  only depends on one 
nonperturbative moment, $A^{(1)}_2$. Moreover, 
in the longitudinal case the coefficient is not enhanced by 
logarithms of $Q_A \, Q_B$. 
For both transverse and longitudinal scattering, 
it would be interesting to use experimental data 
 for  studying   
the moments  $A^{(i)}_2$ and for  testing  the dispersive 
structure of the power-behaved terms. 

\vskip 1 cm 

I greatly benefited from discussions with Yu.\ Dokshitzer, 
D.\ Soper and B.\ Webber. I  thank the High Energy Theory group
at Brookhaven National Laboratory, the RIKEN-BNL Research Center 
and the organizers of the RIKEN Workshop on Perturbative QCD 
for hospitality and support while part of this work was being done. 
This research was partially funded  by the US Department of
Energy grant DE-FG03-96ER40969.

%\clearpage
%%
%\count88 = 0
%%
%\begin{figure}[htb]
%\centerline{ \desepsf(loglog.eps width 15 cm) }
%\bigskip
%\caption{$Q^2$-behavior of  .}
%\label{fig:loglog}
%\end{figure}

\end{document}